\documentclass[final]{aipproc}

\layoutstyle{8x11single}

\usepackage{amsmath}
\usepackage{amsfonts}
\usepackage{mathrsfs}
\usepackage{slashed}

\begin{document}
\title{Baryon and lepton number violating effective operators in a non-universal extension of the Standard Model}

\classification{11.30.Fs, 11.30.Hv, 12.60.Cn}
\keywords      {Baryon Number Violation, Lepton Number Violation, Non-perturbative effects}

\author{J.~Fuentes-Mart\'in}{
  address={Instituto de F\'{\i}sica Corpuscular, CSIC - Universitat de Val\`encia, \\
Apt. Correus 22085, E-46071 Val\`encia, Spain}
}

\begin{abstract}
It is well known that non-abelian Yang-Mills theories present non-trivial minima of the action, the so-called instantons. In the context of electroweak theories these instanton solutions may induce violations of baryon and lepton number of the form $\Delta B = \Delta L = n_f$, with $n_f$ being the number of families coupled to the gauge group. An interesting feature of these violations is that the flavor structure of the gauge couplings is inherited by the instanton transitions. This effect is generally neglected in the literature. We will show that the inclusion of flavor interactions in the instanton solutions may be interesting in certain theoretical frameworks and will provide an approach to include these effects. In particular we will perform this implementation in the non-universal $SU(2)_l \otimes SU(2)_h \otimes U(1)_Y$ model that singularizes the third family. Within this framework, we will use the instanton transitions to set a bound on the $SU(2)_h$ gauge coupling.
\end{abstract}

\maketitle

\section{Introduction}
It has been known for some time that baryon ($B$) and lepton ($L$) numbers are violated in the electroweak sector of the Standard Model (SM) due to anomalies~\cite{Belavin:1975fg,Polyakov:1975rs}. This violation takes place in such a way that, at lowest order, $\Delta B=\Delta L=n_f$ with $n_f$ the number of families coupled to the gauge group and therefore the quantity $B-L$ remains conserved. 't Hooft realized that the explicit violation of these global symmetries is due to classical gauge configurations with non-trivial topological charge~\cite{'tHooft:1976up,'tHooft:1976fv}. These gauge configurations are termed instantons and describe tunneling transitions between different inequivalent vacua. At zero temperature the potential barrier that separate the different vacua has a huge height, which gives rise to a suppression factor ${\cal O}\left( \exp \left[ - 8 \pi^2/g^2 \right] \right)$ for these $B+L$ violating processes. It has been suggested that $B+L$ violating processes might be unsuppressed in high-energy collisions~\cite{Arnold:1987zg} where the vacuum transitions, denoted now sphalerons, take place from above the potential barrier and therefore are free of the exponential suppression. The computation of these processes was done in Refs.~\cite{Espinosa:1989qn,Ringwald:1989ee}. Unfortunately, the calculations performed in this direction violate the unitarity bound and therefore are unreliable.

Even though the SM is in perfect agreement with the current experimental data, several theoretical and experimental issues need to be addressed. They have been extensively treated in the literature giving rise to many theories Beyond the Standard Model (BSM). Although low-energy instanton transitions are highly suppressed in the SM model this might no longer be true for BSM theories where the gauge couplings are larger. This possibility was explored in the framework of gauge non-universal models in Ref.~\cite{Morrissey:2005uza} where the inclusion of flavor dynamics was missing in the calculation. This talk follows closely the work done in Ref.~\cite{Fuentes-Martin:2014fxa} and is devoted to the introduction of inter-family mixing in the one-instanton transitions. We will show that these effects are crucial in the calculation of proton decay observables and will present a systematic approach to its inclusion in the model. Once that the instanton-mediated baryon and lepton number violating amplitudes have been calculated, we will obtain an effective Lagrangian for these interactions that violates not only baryon and lepton number but also flavor. These operators will be used in order to constrain the gauge couplings from proton decay. For the sake of concreteness, we will perform this calculation in the non-universal $SU(2)_l \otimes SU(2)_h \otimes U(1)_Y$ model. However, many of the results presented here can be easily applied to other new physics models. 

\section{The non-universal $\mathbf{SU(2)_{l}\otimes SU(2)_h \otimes U(1)_Y}$ model}
We will analyze non-perturbative processes of an electroweak extension of the SM given by the symmetry group $\mathcal{G}\equiv SU(2)_{l} \otimes SU(2)_h \otimes U(1)_Y$ that breaks family universality by singularizing one of the families, in this case the third one~\cite{Li:1981nk,Muller:1996dj}. This extension embeds the SM gauge group and provide a good description of the experimental data together with some interesting phenomenological predictions~\cite{Chiang:2009kb,Hsieh:2010zr,Kim:2011qk,Jezo:2012rm,Cao:2012ng,Kim:2014afa,Edelhauser:2014yra}. The fermion content of this model is the same as in the SM:
\begin{align} \label{eq:frepr}
\begin{aligned}
Q_{i}&:\;\left(2,1\right)\left(1/3\right), &\qquad Q_{3}&:\;\left(1,2\right)\left(1/3\right),  \\
L_{i}&:\;\left(2,1\right)\left(-1\right), &\qquad L_{3}&:\;\left(1,2\right)\left(-1\right),  \\
u_{j}&:\;\left(1,1\right)\left(4/3\right), &\qquad d_{j}&:\;\left(1,1\right)\left(-2/3\right),  \\
e_{j}&:\;\left(1,1\right)\left(-2\right), &\qquad  
\end{aligned}
\end{align}
with the first parenthesis showing the group representation under $SU(2)_{l} \otimes SU(2)_h$ while the second parenthesis indicates the hypercharge. On the other hand, the model requires the introduction of two Higgs doublets, $\Phi_l$ and $\Phi_h$, to generate the fermion masses of the first two families and the third one, respectively. Additionally, it is necessary to include a bi-doublet in order to recover the SM gauge group at low energies via spontaneous symmetry breaking. This way, the symmetry transformations for scalar sector read:
\begin{align} \label{eq:hrepr}
\begin{aligned}
\Phi_l&:\;\left(2,1\right)\left(1\right), &\qquad \Phi_h&:\;\left(1,2\right)\left(1\right),  \\
\Phi_b&:\;\left(2,2\right)(0). & 
\end{aligned}
\end{align}
The bi-doublet has to be self-dual under the $SU(2)$ group so that the correct symmetry breaking pattern is triggered, that is, it should satisfy the condition $b=\tau^2 b^*\tau^2$ with $\tau^2$ the Pauli matrix. The bi-doublet will then acquire a vacuum expectation value (vev) $\langle \Phi_b \rangle = u/2 \; \mathbb{I}$, with $u$ of order of few TeVs which gives rise to the spontaneous symmetry breaking pattern $\mathcal{G}\to SU(2)_L \otimes U(1)_Y$. In a second step the Higgs doublets will also acquire a vev $\langle \Phi^0_{l,\,h}\rangle = v_{l,\,h}/\sqrt{2}$ with $v=\sqrt{v_l^2+v_h^2}\simeq 0.246 \; \mbox{TeV}$ that breaks the SM gauge group down to electromagnetism.

The Lagrangian for this theory is given by:
\begin{align} \label{eq:lagrange}
\begin{aligned}
{\cal L}  =&  - \frac{1}{4} \sum_{a=1}^3 W_{l \, \mu \nu}^a W_{l}^{a \, \mu \nu} \, - \, \frac{1}{4} \sum_{a=1}^3 
W_{h \, \mu \nu }^a W_h^{a \, \mu \nu} \, - \, \frac{1}{4} B_{\mu \nu} B^{\mu \nu} \, \\
& + \, \sum_{i=1}^3  i \, \psi_{L\,i}^\dagger  \, \overline{\sigma}^{\mu} \,  D_{\mu} \,  \psi_{L\,i} \, 
+ \, \sum_{j=1}^3  i \, \psi_{R\,j}^\dagger \,  \sigma^{\mu} \,  D_{\mu} \,  \psi_{R\,j} \, + \, {\cal L}_{\mbox{\tiny{Y}}}\\
& + \, \sum_{r=h,l} \, 
\left( D_{\mu} \Phi_r \right)^\dagger \left( D^{\mu} \Phi_r \right) \, + \, \mbox{Tr} \left[ \left(D_{\mu} \Phi_b \right)^\dagger \left( D^{\mu} \Phi_b \right) \right] \,  \, - \, V\left[\Phi_s \right] ,
\end{aligned}
\end{align}
where we have used a compact notation for the scalars, $\Phi_s$ with $s=l,\,h,\,b$, and for the fermions where $\psi$ stands for both quark and leptons with $i$ and $j$ denoting flavor. The spin matrices are defined as $\sigma^{\mu} = \left(1, \vec \sigma \right)$ and $\overline{\sigma}^{\mu} = \left( 1, - \vec \sigma \right)$ and $W_{\mu\nu}^a$ and $B_{\mu\nu}$ are the field-strength tensors for the $SU(2)$ and the $U(1)_Y$ gauge groups, respectively. The covariant derivatives are defined as:
\begin{align} \label{eq:dcov}
\begin{aligned}
D^{\mu} \, \psi_{L \, i} \, =& \, \left( \partial^{\mu} - i g_h  W_h^{\mu} \, \delta_{i 3} -i g_{l}  W_{l}^{\mu} \, \delta_{i \{1,2\}} - i \frac{g'}{2} Y B^{\mu} \right) \, \psi_{L \, i} , \\
D^{\mu} \, \psi_{R \, j} \, =& \, \left( \partial^{\mu} - i g' Q B^{\mu} \right) \, \psi_{R \, j}, \\
D^{\mu} \, \Phi_r \, =& \, \left( \partial^{\mu} - i g_h  W_h^{\mu} \, \delta_{r h} -i g_{l}  W_{l}^{\mu} \, \delta_{r l} - i \frac{g'}{2}  Y  B^{\mu} \right) \, \Phi_r , \; \; \; \; r=l,h, \\[0.3cm]
D^{\mu} \, \Phi_b \, =& \,  \partial^{\mu} \, \Phi_b \, + \,  i g_h \, W_h^{\mu} \, \Phi_b - \, i g_{l} \,  W_{l}^{\mu} \, \Phi_b .
\end{aligned}
\end{align}
Here $Q$ stands for the electric charge defined as $Q= \tau^3/2 + Y/2$ and $Y$ denotes the hypercharge, which is given in the second parenthesis of Eqs.~\eqref{eq:frepr} and~\eqref{eq:hrepr}. The Yukawa Lagrangian is given by:
\begin{align} \label{eq:yukawa}
\begin{aligned}
- {\cal L}_{\mbox{\tiny{Y}}} = &\;  Y_{ij}^l \, u^{\dagger}_{i} \, \tilde{\Phi}_l^{\dagger} \, Q_{j} \, + \, 
Y_{i3}^h \,  u^{\dagger}_{i} \, \tilde{\Phi}^{\dagger}_h \, Q_{3} \, + \, 
X_{ij}^l \, d^{\dagger}_{i} \, \Phi_l^{\dagger} \, Q_{j} \, + \, 
X_{i3}^h \, d^{\dagger}_{i} \, \Phi^{\dagger}_h \, Q_{3} \, \\
& + \, Z_{ij}^l \, e^{\dagger}_{i} \, \Phi_l^{\dagger} \, L_{j} \, +
\, Z_{i3}^h \, e^{\dagger}_{i} \, \Phi_h^{\dagger} \, L_{3} \, + \, h.c. ,
\end{aligned}
\end{align}
where $i = 1,\,2,\,3$ and $j=1,\,2$ denote family and $\tilde{\Phi} \equiv \varepsilon \Phi^{*} = i \tau^2 \Phi^{*}$. For the issues considered here and without lack of generality, we will assume that the Yukawa couplings $X^{l,\,h}$, $Y^{l,\,h}$ and $Z^{l,\,h}$ are real.

Finally, the gauge couplings of the $SU(2)_l$ and $SU(2)_h$ gauge groups are related to the $SU(2)_L$ coupling according to the relation
\begin{align}
g=\frac{g_l\;g_h}{\sqrt{g_l^2+g_h^2}}\,,
\end{align}
and therefore $g_l$ and $g_h$ are always bigger than the SM coupling and are unbounded from above. This allows us to work in the limit where one of the couplings is large while the other approaches the SM value. In what follows, we will assume that $g_h$ is larger that $g_l$ and, accordingly, we will only focus on the computation regarding the $SU(2)_h$ instantons.

\section{Instanton-mediated amplitudes}
In this section we will calculate the instanton-mediated Green functions generated by vacuum transitions in the background of the one-instanton solution that minimizes the Euclidean action. Although, strictly speaking, the Green functions are computed in Euclidean space the results we show in this section correspond to an analytic continuation from these Green functions to Minkowsky space. We will see that, when flavor violating gauge interactions are included in the computation, the instanton-generated Green functions not only violate lepton and baryon number but also flavor. In a second step we will make use of the LSZ formula to compute the corresponding amplitudes from the Green functions. As we anticipated, we will work in the limit where $g_h$ is larger than $g_l$ (but still perturbative) and we only consider the instanton background generated by the $SU(2)_h$ gauge group and so in order not to stress the notation we will drop the $h$ subindex in $W_h$ in what follows. These Green functions are written as:
\begin{align}\label{eq:GF}
G\left(x_1,\dots,x_{\mbox{\tiny $N_f$}}\right)=\langle\prod_{i=1}^{\mbox{\tiny $N_f$}}q_i\left(x_i\right)\rangle_I \,,
\end{align}
with $N_f$ the number of fermions coupled to the gauge group, in this case $N_f=4$.

One difficulty appears when the scalar fields develop a vev as, for $\langle\Phi_s^0\rangle\neq0$, the action has no non-trivial stationary points. However an approximate instanton solution can still be found in the region where $\rho\langle\Phi_s^0\rangle$, being $\rho$ the instanton radius, is small, with an exponential fall-off outside this region. This approximate solution, which was anticipated by 't Hooft~\cite{'tHooft:1976fv}, was formally developed by Affleck with the introduction of the \textit{constrained instanton formalism}~\cite{Affleck:1980mp}. Under this formalism the instanton solution in the \textit{singular gauge} for both the short-distance $\left(x\ll\rho\right)$ and the long-distance $\left(x\gg\rho\right)$ regimes reads:
\begin{align}
\label{eq:AI}
g_h \, W_{\mu,\,I}^a\left(x\right)  \, \frac{\tau^a}{2} &=  \mathscr{W}_I(x)\, x_{\nu} \,  \overline{\eta}_{\mu\nu}^a \, U \, \tau^a \, U^\dagger \,,
\end{align}
where $U$ parameterizes the instanton gauge orientation and the leading order expression for $\mathscr{W}_I$ is given by:
\begin{align}\label{eq:Arara}
\mathscr{W}_I(x)= \begin{cases} 
                             \; \;   \rho^2 \,  \frac{1}{x^2 \, \left(x^2+\rho^2\right)}  \,, & x\ll\rho , \\[0.6cm]
                             \; \;    \rho^2 \, M_W^2 \, \frac{K_2\left(M_Wx\right)}{2 \, x^2}  \,, & x\gg\rho , \\
                                \end{cases}
\end{align}
with $M_W$ being the mass of the gauge boson after spontaneous symmetry breaking and $K_2(x)$ a Modified Bessel function of the second kind. The corresponding anti-instanton solution can be obtained by the replacement $\overline{W}^\mu_I = W^\mu_I \left( \overline{\eta}_{\mu \nu}^a \longrightarrow \eta_{\mu \nu}^a \right)$, where the symbols $\eta_{\mu\nu}^a$ and $\overline{\eta}_{\mu\nu}^a$ relate the $SU(2)$ and the $SO(4)$ generators (see Ref.~\cite{'tHooft:1976fv}). 

Regarding the instanton solution for the Higgs field, it is convenient to work on the Higgs basis where only the Higgs doublet that acquires a vev has a non-zero instanton solution. The leading order solution reads:
\begin{align}
\label{eq:inshig}
 \Phi_I \left(x\right) & =   c_\beta \, \Phi_{l,\,I}(x) \, + \, s_\beta \, \Phi_{h,\,I}(x)  = 
  \begin{cases}
    \left[ c_\beta \, \langle \Phi_l^0 \rangle \, + \, s_\beta \, \left(\frac{x^2}{x^2+\rho^2}\right)^{\frac{1}{2}}   \langle \Phi_h^0 \rangle \right] \, \overline{h}, &  x\ll \rho , \\[0.6cm]
 \; \left[ c_\beta \, \langle \Phi_l^0 \rangle  \, + \, s_\beta \, \langle \Phi_h^0 \rangle \, \right] \, \overline{h}\;= \frac{v}{\sqrt{2}}\,\overline{h} , & x\gg \rho ,
\\
 \end{cases}
\end{align}
where $\tan\beta = \langle \Phi_h^0 \rangle/\langle \Phi_l^0 \rangle$ and $\overline{h}=\left(0,1\right)^T$ is a constant isospinor.

The general approach to compute the instanton-mediated Green functions is based on a perturbative semi-classical expansion of the Euclidean action around the classical instanton configuration up to one-loop
\begin{align}\label{eq:bfmf}
\begin{aligned}
W_\mu^a &= W_{\mu,\,I}^a+\delta W_\mu^a\,, \\
\Phi_s &= \Phi_{s,\,I}+\delta\Phi_s\,,
\end{aligned}
\end{align}
where $\delta$ denotes a quantum fluctuation and the rest of the fields remain at the quantum level.

Under this expansion the one-instanton Green functions in momentum space take the general form:
\begin{align}\label{eq:GF2}
\widetilde{G}\left(p_1,\dots, p_{N_f}\right)= \left(2\pi\right)^4\delta^4\left(\sum_{i=1}^{N_f} p_i\right) \, \int dU\int d\rho\,e^{-S^{\mbox{\tiny cl}}_{\mbox{\tiny E}}\left[ W_I, \Phi_{s,\,I} \right]}
F\left(\rho;\mu\right)\prod_{i=1}^{N_f} \widetilde{\psi}_{0,\,i}\left(p_i\right)\,,
\end{align}
where $\rho$ and $U$ are {\em collective coordinates} parameterizing instanton size and gauge orientation. The function $F\left(\rho;\mu\right)$ was calculated in Ref.~\cite{'tHooft:1976fv}:
\begin{align}
\label{eq:frhomu}
F\left(\rho;\mu\right) \, = \, C \, g_h^{-8} \, \left(\rho\mu\right)^{\beta_1}\, \rho^{-5}\,,
\end{align}
with $\mu$ being the normalization point in the $\overline{\mbox{MS}}$ scheme and the factor $C$ given by:
\begin{align}
\label{eq:cthooft}
C=2^{10}\pi^6e^{-\alpha\left(1\right)+\left(N_f-N_S\right)\alpha\left(\frac{1}{2}\right)+\frac{5}{36}\left(2-\frac{1}{2}N_f+\frac{1}{2}N_S\right)}.
\end{align}
Here $\alpha(1)\simeq0.443$ and $\alpha\left(\frac{1}{2}\right)\simeq0.146$ and $\beta_1$ is the one-loop beta function of $SU(2)_h$:
\begin{align}\label{eq:onelbeta}
\beta_1=\frac{22}{3}-\frac{1}{3}N_f-\frac{1}{6}N_S\,,
\end{align}
where, we remind, $N_f=4$ and $N_S=2$ is the number of $SU(2)_h$ scalar doublets (with the bi-doublet counting as one).

The classical action contain only classical fields and it gives:
\begin{equation}
\label{eq:SCL}
S^{\mbox{\tiny cl}}_{\mbox{\tiny E}}\left[W_I,\Phi_{s,\,I}\right] \, \simeq \, \frac{8\pi^2}{g_h^2} \, + \, 4 \, \pi^2 \rho^2 \, \, \mathcal{V}^2,
\end{equation}
with $\mathcal{V}^2 = \sum_i q_i \langle \Phi_i^0 \rangle^2 = 1/4 \left[ v_h^2 + u^2 \right] \simeq u^2/4$. The second term in the above equation is only present in theories with spontaneous symmetry breaking and guarantees that the Green functions are infrared safe, a feature that was realized by 't Hooft~\cite{'tHooft:1976fv}. Finally, $\widetilde{\psi}_{0,\,i}\left(p_i\right)$ denote the instanton zero modes, which are the normalizable solutions of the fermion operator. These introduce fermion number violation in the theory and its value depend on the model. We will deal with its computation in the next section.

\subsection{Fermion zero modes}
In this section we will compute the fermion zero modes associated to the $SU(2)_h$ gauge group. It is in this part where the non-trivial flavor dynamics takes places. For simplicity, we will perform the calculation only in the quark sector as the generalization for the lepton sector is straightforward. As instantons are solutions of the Euclidean action we need to work in Euclidean space in order to calculate the fermion zero modes. Before presenting the Lagrangian of the model in Euclidean form, one technicality should be taken into account: the fermion spin representations that in the Minkowsky space belong to the $SO(3,1)$ group need to be transformed into representations of the $SO(4)$ group where the two spinor representations are no longer related by complex conjugation. The relation between the $SO(3,1)$ representations ($\psi_{L,\,R}$) and the ones of $SO(4)$ ($\chi_{A,\,B}$) is given by:
\begin{align} \label{eq:eucl}
\psi_R \, \rightarrow \, \chi_A, \;\;\;\; \;  \; \;  \psi_L \, \rightarrow \, \chi_B ,\;\; \; \; \; \;\; \psi_R^\dagger \, \rightarrow \, \chi_B^\dagger, \;\;\; \; \; \; \;   \psi_L^\dagger \, \rightarrow \, \chi_A^\dagger .
\end{align}

Additionally, it is convenient to perform the computation in the so-called Higgs basis for the Higgs doublets and to use a flavor diagonal basis for the fermions. The latter is achieved by performing the following unitary transformation:
\begin{align}\label{eq:mass-basis}
\begin{aligned}
u_i\to\left(V_u^\dagger\right)_{ij}u_j,  \\
d_i\to\left(V_d^\dagger\right)_{ij}d_j, 
\end{aligned}
\end{align}
with $i$ and $j$ being flavor indices and the unitary flavor matrices defined such that they diagonalize the mass matrices.

With all this into consideration and keeping only the relevant operators for the current calculation, the Euclidean Lagrangian of the model reads (note that the fermion fields are now different than in Eq.~\eqref{eq:lagrange} and related to those through the unitary transformation defined in Eq.~\eqref{eq:mass-basis}):
\begin{align} \label{eq:lagrangee}
\begin{aligned}
{\cal L}_{\mbox{\tiny E}}  =& \;  \frac{1}{4} \sum_{a=1}^3 W_{h \, \mu \nu }^a W_{h \, \mu \nu}^{a} \,+ \, \left( D_\mu \Phi_h \right)^\dagger \left( D_\mu \Phi_h \right) \,+ \, \left( \partial_\mu \Phi_l \right)^\dagger \left( \partial_\mu \Phi_l \right) \,+ \, \mbox{Tr} \left[ \left(D_\mu \Phi_b \right)^\dagger \left( D_\mu \Phi_b \right) \right] \,  + \, V\left[\Phi_s \right]  \\
&    +  \,  i \, Q_{A}^\dagger \,  \hat{\overline{\sigma}}_\mu \,  \tilde{D}_\mu \,  Q_{B} \, +  \,  i \, u_{B}^{\dagger} \,  \hat{\sigma}_\mu \,  \partial_\mu \,  u_{A}  \, +  \,  i \, d_{B}^\dagger \,  \hat{\sigma}_\mu \,  \partial_\mu \,  d_{A}\, + \, \mathcal{L}_Y^{\mbox{\tiny{E}}}\,+\dots \, , 
\end{aligned}
\end{align}
where the sum over families in the kinetic terms is implicit and with the Euclidean spin matrices given by $\hat{\sigma}_{\mu}=-\left(\vec\sigma,i\right)$ and $\hat{\overline{\sigma}}_{\mu}=\left(\vec\sigma,-i\right)$. The covariant derivatives acting on the scalar fields are now:
\begin{align}
\begin{aligned}
D_{\mu} \, \Phi_h \, =& \, \partial_\mu\Phi_h - i g_h  W_\mu \, \Phi_h , \\[0.3cm]
D_{\mu} \, \Phi_b \, =& \,  \partial_\mu \, \Phi_b \, + \,  i g_h \, W_\mu \, \Phi_b\,,
\end{aligned}
\end{align}
and the new covariant derivative appears as a consequence of the flavor rotation and is defined as:
\begin{align}\label{eq:dcovnonu}
\begin{aligned}
\tilde{D}_\mu &=\partial_\mu-\frac{1}{2}i \, g_h \, W_\mu^a \, \mathscr{F}^\dagger \, \tau^a \, \mathscr{F}  \\
&=\partial_\mu-\frac{1}{2}i \, g_h \begin{pmatrix}
                                W_\mu^0 \, P_u & \sqrt{2} W_\mu^+ \, P_u \, V_{\mbox{\tiny CKM}} \, P_d\\
                             \sqrt{2}W_\mu^- \, P_d \, V_{\mbox{\tiny CKM}}^\dagger \, P_u & -W_\mu^0 \,  P_d
                                        \end{pmatrix} \,.
\end{aligned}                                        
\end{align}
Here the matrices $P_{u,\,d}$ project the mass eigenstates into the third family in the gauge diagonal basis. Their expression in terms of the flavor matrices takes the form:
\begin{align}
\begin{aligned}
\left(P_f\right)_{ij}&=\left(V_f\right)_{i3}\left(V_f^\dagger\right)_{3j},
\end{aligned}
\end{align}
where $f=u,\,d$ while the matrix $\mathscr{F}$ is given by:
\begin{align}
 \mathscr{F}=\begin{pmatrix}
             P_u & 0\\
             0 & V_{\mbox{\tiny CKM}}P_d
             \end{pmatrix}=
             \begin{pmatrix}
             P_u & 0\\
             0 & P_u\,V_{\mbox{\tiny CKM}}
             \end{pmatrix}\,,            
\end{align}
with $V_{\mbox{\tiny CKM}}=V_u\,V_d^\dagger$ being the Cabibbo-Kobayashi-Maskawa matrix. Finally, the Yukawa Lagrangian in Euclidean space reads:
\begin{align} \label{eq:yuke}
\begin{aligned}
{\cal L}_Y^{\mbox{\tiny E}} =&   \,  \lambda^u_{ij}\left( u_{B \, i}^\dagger \, \tilde{\Phi}^\dagger \, Q_{B \, j} \, + \, Q_{A \, i}^\dagger \, \tilde{\Phi} \, u_{A \, j}\right)\, + \,  \lambda^d_{ij} \, \left( d_{B \, i}^\dagger \, \Phi^\dagger \, Q_{B \, j}\, + \,  Q_{A \, i}^\dagger \, \Phi \, d_{A \, j}  \right)\,+\dots
\end{aligned}
\end{align}
Here the new matrices are defined as $\lambda^f_{ij}=m_{f_i}/\langle \Phi^0 \rangle\,\delta_{ij}$ with $f=u,\,d$ and $\Phi$ is the doublet that acquires a vev in the Higgs basis while the dots stand for interactions with the other doublet, which are irrelevant for the present calculation.

In order to compute the fermion zero modes we need to solve the equations of motion for the fermion fields in the instanton background, that is for $W_\mu^a = W_{\mu,\,I}^a $ and $\Phi = \Phi_I$ (see Eqs.~\eqref{eq:AI} and~\eqref{eq:inshig}):
\begin{align}\label{eq:EOM}
\begin{aligned}
i \,\hat{\overline{\sigma}}_{\mu} \tilde{D}_\mu \, Q_{B \, i} \, + \, \lambda^u_{ij} \, \varepsilon \, \Phi^*_I \, u_{A \, j} \, + \,\lambda^d_{ij} \, \Phi_I \, d_{A \, j}  = & \; 0, \\
- \, \lambda^u_{ij} \, \Phi^T_I \, \varepsilon \, Q_{B \, j} \, + \, i\hat{\sigma}_\mu  \partial_\mu  u_{A \, i}  = & \; 0, \\
 \lambda^d_{ij}\, \Phi^\dagger_I \,  Q_{B \, j} \, + \, i\hat{\sigma}_\mu  \partial_\mu  d_{A \, i}  = & \; 0\,.
\end{aligned}
\end{align}

The above equations get simplified if we use the following ansatz:
\begin{align} \label{eq:ansatze}
\begin{aligned}
Q_{B \, i}\left(x\right) = & \; x_\mu \, \hat{\sigma}_\mu \, \xi_{B \, i}\left(y\right), \\
u_{A \, i} = & \; u_{A \, i}\left(y\right), \\
d_{A \, i} = & \; d_{A \, i}\left(y\right),
\end{aligned}
\end{align}
with $y=x_\mu x_\mu$. This way the equations of motion now read:
\begin{align}\label{eq:EOM_ansatz}
\begin{aligned}
\mathscr{D}_{\mbox{\tiny U}} \, \xi_{B \, i}(y) \, + \, i \, \lambda^u_{ij} \, \varepsilon \, \Phi_I^* \, u_{A \, j}(y) \, + \, i \, \lambda^d_{ij} \, \Phi_I \, d_{A \, j}(y) = & \; 0, \\
- \lambda^u_{ij} \, \Phi_I^T \, \varepsilon \, \xi_{B \, j}(y) \, + \, 2i \, \left( u_{A \, i}(y) \right)' = & \; 0, \\
\lambda^d_{ij}\, \Phi_I^\dagger \, \xi_{B \, j}(y) \, + \, 2i \, \left( d_{A \, i}(y) \right)' = & \; 0,
\end{aligned}
\end{align}
where we have defined a new derivative
\begin{align} \label{eq:dmuchi}
\mathscr{D}_{\mbox{\tiny U}} \, \xi_{B \, i}(y) \, = \, 4 \, \xi_{B \, i}(y) \, + \, 2 \, y \, \left(\xi_{B \, i}(y)\right)' \, + \, y \, \mathscr{W}_I(y) \, \mathscr{F}^\dagger \, U \, \left(\vec{\sigma} \, \cdot \, \vec{\tau}\right) \, U^{\dagger} \, \mathscr{F}  \, \xi_{B \, i}(y)\,,
\end{align}
with $\left(\vec{\sigma} \, \cdot \, \vec{\tau}\right)$ belonging to the coupled spin-isospin space. From Eqs.~\eqref{eq:EOM_ansatz} and~\eqref{eq:dmuchi} it is clear that the flavor structure of the zero modes is completely determined by the last term in the above equation. Moreover, this structure is the same as in the covariant derivative in the mass-diagonal basis.

A solution to Eqs.~\eqref{eq:EOM_ansatz} can be obtained in both the short-distance and the long-distance regimes. However, as shown in Refs.~\cite{Espinosa:1989qn,Ringwald:1989ee},
only the long-distance solution contributes to the instanton-mediated Green functions. The reason for that is that only the long-distance solutions in momentum space present poles in $p=\pm im$ and, in accordance to the LSZ procedure, only the pole part of the Green functions contributes to the amplitudes. Unfortunately, the leading order long-distance solutions are flavor blind and therefore the flavor factor in the long-distance regime is undetermined. This is because in the long-distance regime the instanton gauge configuration satisfies $y\mathscr{W}_I\to0$. The general approach to obtain the flavor structure consists then in solving Eqs.~\eqref{eq:EOM_ansatz} in the short-distance regime and then match the solutions in the intermediate region to fix the global factors. It is enough to obtain the solution just to the first equation in Eqs.~\eqref{eq:EOM_ansatz}, which at first order is given by:
\begin{align}
\xi_{B \, i}\left(x\right)  = & \;\frac{\sqrt{2}}{\pi}\frac{\rho^{3/2}}{x\left(x^2+\rho^2\right)^{3/2}} \, \mathscr{F}^\dagger \, U \, \zeta_{s_i}\,,
\end{align}
with $\zeta_{s_i} = (0,1,-1,0)^T/\sqrt{2}$ and orthogonal in flavor space. The global factor, other than the flavor structure denoted by $\mathscr{F}$, has been chosen so that the zero modes are correctly normalized. This solution in turn fixes the flavor factors of the long-distance solutions which take the form (for more details on their derivation see Ref.~\cite{Espinosa:1989qn}):
\begin{align} \label{eq:ldsol}
\begin{aligned}
f_{A \, j}\left(x\right) & = \frac{i}{2\pi} \, \rho^{3/2} \,  m_{f_j}^2\, \frac{K_1\left(m_{f_j} \,x\right)}{x}U \, \chi_f^P , \\
f_{B \, j}\left(x\right) & = -\frac{1}{2\pi} \, \rho^{3/2} \,  m_{f_j}^2\, \frac{K_2\left(m_{f_j} \, x\right)}{x^2} \, x_\mu \hat{\sigma}_\mu \,  U \, 
\chi_f^P  , 
\end{aligned}
\end{align}
where $f=u,\,d$ and the normalized projected spinors are defined as:
\begin{align}
\begin{aligned}
\chi_u^P&\equiv\left(V_u\right)_{i3}\chi_{u_i},\\
\chi_d^P&\equiv \left( V_d \right)_{i3} \, \chi_{d_i} \, = \, \left(V_{\mbox{\tiny CKM}}^\dagger\right)_{im}\left(V_u\right)_{m3}\chi_{d_i}.
\end{aligned}
\end{align}
Here $\chi_{u_j} = (0,1)^T$ and $\chi_{d_j} = (-1,0)^T$ are spinors orthogonal in flavor space.

Finally, performing a Fourier transformation to the solutions in Eq.~\eqref{eq:ldsol}, reverting to Minkowsky space and assembling the Weyl spinors into a Dirac spinor in the Weyl basis, we obtain the following amputated fermion zero modes in momentum space (see Refs.~\cite{Espinosa:1989qn,Fuentes-Martin:2014fxa} for more details):
\begin{align} \label{eq:ftamp}
\begin{aligned}
\left[u(p)\right]_{\mbox{\footnotesize Amp}}&=-2\pi i \, \rho^{3/2} \, \left(\begin{matrix}
                                                              0\\
                                                              U \, \chi_u^P
                                                              \end{matrix}\right) , \\
\left[d(p)\right]_{\mbox{\footnotesize Amp}}&=-2\pi i \, \rho^{3/2} \, \left(\begin{matrix}
                                                                 0\\
                                                                 U \, \chi_d^P
                                                                 \end{matrix}\right).
\end{aligned}                                                                 
\end{align}
The zero modes in the lepton sector can be obtained just by substituting in the above equation $u\to\nu$, $d\to e$ and $V_{\mbox{\tiny CKM}}\to I$ (or, if we consider neutrinos masses, $V_{\mbox{\tiny CKM}}\to U_{\mbox{\tiny PMNS}}$ with $U_{\mbox{\tiny PMNS}}$ the Pontecorvo-Maki-Nakagawa-Sakata matrix). 

\subsection{Integration over collective coordinates}
Once that we have calculated the amputated fermion zero modes we can invoke the LSZ procedure to obtain from the Green functions in Eq.~\eqref{eq:GF2} the instanton-mediated baryon and lepton number violating amplitudes. These take the general form:
\begin{align}\label{eq:amplitude1}
A=& \, C \, g_h^{-8} \, e^{-\frac{8\pi^2}{g_h^2\left(\mu\right)}} \, \left(2\pi i \right)^{N_f}
\int d\rho\,e^{-4\pi^2\rho^2\mathcal{V}^2} \, \rho^{\frac{3N_f}{2}-5} \, \left( \mu \rho \right)^{\beta_1} \, \int dU \, \prod_{f=1}^{N_f}
\overline{\omega_{f}} \, \, \left(\begin{matrix}
                                                                 0\\
                                                                 U \chi_{f}^P
                                                                 \end{matrix}\right),
\end{align}
where $\omega_{f}$ with $f=u,\,d,\,e,\,\nu$ is an external-state polarization spinor and, we remind, $\mathcal{V}^2=1/4 \left[ v_h^2 + u^2 \right] \simeq u^2/4$, $N_f=4$ and $N_S=2$ for the model we are considering. In order to compute the amplitude only the calculation of the integrals in instanton gauge orientation and instanton size remain. The latter is easily performed and gives:
\begin{align} \label{eq:igamma}
\int_0^{\infty} d\rho\,e^{-4\pi^2\rho^2\mathcal{V}^2}\rho^{\frac{3N_f}{2}-5+\beta_1}=\frac{1}{2}\left(\frac{1}{4\pi^2\mathcal{V}^2}\right)^{\frac{3N_f}{4}+\frac{\beta_1}{2}-2}\Gamma\left(\frac{3N_f}{4}+\frac{\beta_1}{2}-2\right).
\end{align}
The integral over gauge orientation is a bit more involved (see Refs.~\cite{Morrissey:2005uza,Fuentes-Martin:2014fxa} for more details). We show here the result:
\begin{align} \label{eq:operatore}
\begin{aligned}
\int dU\prod_{f=1}^{4}\overline{\omega_f} \,\left(\begin{matrix}
                                                                 0\\
                                                                 U \, \chi_f^P
                                                                 \end{matrix}\right) \, = & \; 
\frac{1}{6} \, \epsilon_{\alpha\beta\gamma} \,\overline{\omega_{u_i}^\alpha} \,\left(V_u\right)_{i3}\left(V_{\mbox{\tiny CKM}}^\dagger\right)_{jm}\left(V_u\right)_{m3} \left(\omega_{d_j}^\beta\right)^C  \\
& \times \left[\overline{\omega_{u_k}^\gamma}\left(V_u\right)_{k3}\left(V_\ell\right)_{l3}
\, \omega_{e_l}^C - \, \overline{\omega_{d_k}^\gamma} \left(V_{\mbox{\tiny CKM}}^\dagger\right)_{kn}\left(V_u\right)_{n3}\, \left(V_\ell\right)_{l3} \omega_{\nu_l}^C \right] \, ,
\end{aligned}
\end{align}
where $i,j,\,\dots$ are family indices and $\alpha, \beta$ and $\gamma$ are color indices. As expected, we have obtained non-perturbative amplitudes which violate baryon and lepton number in one unity. These amplitudes can be encoded into a set of dimension-six effective operators. This is done in the following section.

\section{Instanton-mediated effective operators and proton decay}
The expression in Eq.~\eqref{eq:amplitude1} after integration over collective coordinates can be seen as the final result. However, it is interesting to reexpress this result in terms of a set of effective operators. The baryon and lepton number violating amplitudes we have calculated in the previous section are mimicked by the following effective Lagrangian:
\begin{align} \label{eq:lblcano}
\mathcal{L}_{B+L}=\left(C_{LL}^e\right)_{ijkl}\left(\mathcal{O}_{LL}^e\right)_{ijkl}+\left(C_{LL}^\nu\right)_{ijkl}\left(\mathcal{O}_{LL}^\nu\right)_{ijkl}+h.c.,
\end{align}
with
\begin{subequations} \label{eq:operatora}
\begin{align}\label{eq:operatorae}
\left(\mathcal{O}_{LL}^e\right)_{ijkl}&=\epsilon_{\alpha\beta\gamma}\overline{\left(u_{L \, i}^{\alpha}\right)^C} \, d_{L\, j}^{\beta} \, 
\overline{\left(u_{L\, k}^{\gamma}\right)^C} \, e_{L\, l},  \\\label{eq:operatoranu}
\left(\mathcal{O}_{LL}^\nu\right)_{ijkl}&=\epsilon_{\alpha\beta\gamma}\overline{\left(u_{L\, i}^{\alpha}\right)^C} \, d_{L \, j}^{\beta} \, 
\overline{\left(d_{L \, k}^{\gamma}\right)^C} \, \nu_{L\, l},
\end{align}
\end{subequations}
and where the Wilson coefficients are defined as:
\begin{subequations}\label{eq:wilsono}
\begin{align}\label{eq:wilsonoe}
\left(C_{LL}^e\right)_{ijkl}&=\frac{C}{12 \, g_h^{8}}\, \mu^{\beta_1} \, e^{-\frac{8\pi^2}{g_h\left(\mu\right)^2}}\, \left(2\pi\right)^{2- \beta_1} \, \left(\frac{1}{\mathcal{V}^2}\right)^{1+\frac{\beta_1}{2}}\, \Gamma\left(1+\frac{\beta_1}{2}\right) \, \left(V^e\right)_{ijkl},\\\label{eq:wilsononu}
\left(C_{LL}^\nu\right)_{ijkl}&=- \, \frac{C}{12 \, g_h^{8}} \, \mu^{\beta_1}e^{-\frac{8\pi^2}{g_h\left(\mu\right)^2}} \, \left({2\pi}\right)^{2-\beta_1} \, \left(\frac{1}{\mathcal{V}^2}\right)^{1+\frac{\beta_1}{2}} \, \Gamma\left(1+\frac{\beta_1}{2}\right) \, \left(V^\nu\right)_{ijkl}.
\end{align}
\end{subequations}
Finally the flavor factor is given by:
\begin{subequations}\label{eq:vpqrs}
\begin{align}
\label{eq:vpqrse}
\left(V^e\right)_{ijkl} &= \left(V_u\right)_{i3}\left(V_{\mbox{\tiny CKM}}^\dagger\right)_{jm}\left(V_u\right)_{m3} \left(V_u\right)_{k3}\left(V_\ell\right)_{l3},\\[0.2cm]
\label{eq:vpqrsnu}
\left(V^\nu\right)_{ijkl} &= \left(V_u\right)_{i3}\left(V_{\mbox{\tiny CKM}}^\dagger\right)_{jm}\left(V_u\right)_{m3}\left(V_{\mbox{\tiny CKM}}^\dagger\right)_{kn}\left(V_u\right)_{n3}\, \left(V_\ell\right)_{l3}. 
\end{align}
\end{subequations}

\begin{figure}[t]
\centering
\includegraphics[width=0.6\textwidth]{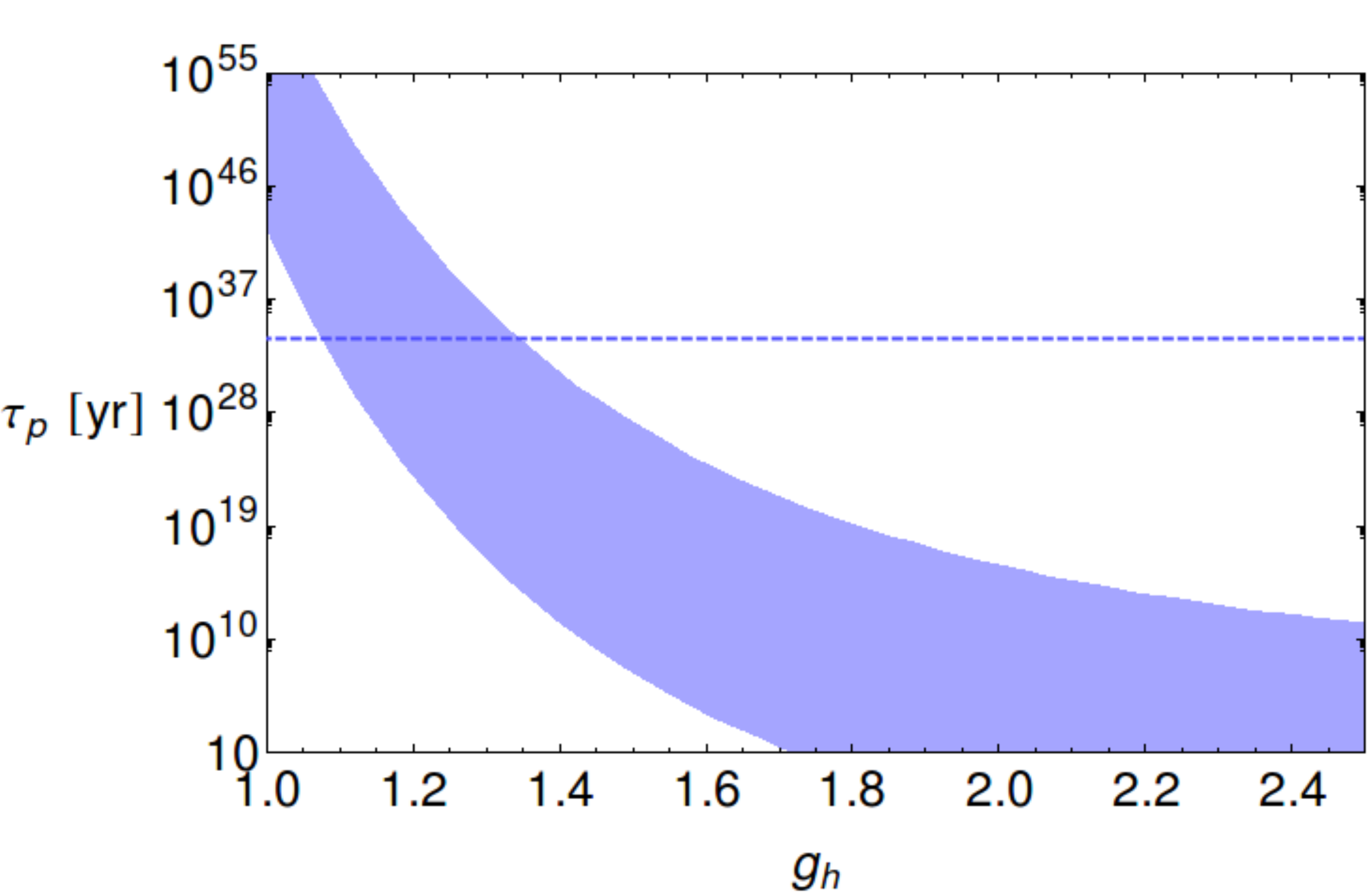}
\caption{Bounds on $g_h$ from the proton decay channel $p\to e^+\pi^0$. The dashed line represents the experimental bound on $p\to e^+\pi^0$ from the PDG~\cite{Beringer:1900zz}, $\tau_{p\to e^+\pi^0}> 8200\times 10^{30}$ years. The band shows the result obtained from the instanton-mediated effective Lagrangian with $\mu=u=3$ TeV and $V_{1111}^e$ in Eq.~\eqref{eq:vpqrse} varying from $10^{-5}$ to $1$.}
\label{fig:protondec}
\end{figure}

The effective Lagrangian introduced in Eq.~\eqref{eq:lblcano} can be used to compute proton decay processes~\cite{Fuentes-Martin:2014fxa}. This way we can obtain a bound on $g_h$ for a fixed value of $u$. The most constrained proton decay channel is given by $p\to e^+\pi^0$, with $\tau_{p\to e^+\pi^0}> 8200\times 10^{30}$ years~\cite{Beringer:1900zz}.  The $B+L$ dimension-six effective Lagrangian gives the following decay width for this channel~\cite{Nath:2006ut}:
\begin{align}\label{eq:pdecay}
\begin{aligned}
\Gamma(p\to e^+\pi^0) & = \frac{(m_p^2-m_\pi^2)^2}{128\pi f_\pi^2\, m_p^3}\,
 \big| \beta (C_{LL}^e)_{1111} \big|^2 \, (1+D+F)^2 \\[2mm]
& \simeq (1.9\cdot 10^{-4}~\mbox{GeV}^5) \, \big|  (C_{LL}^e)_{1111} \big|^2
\,.
\end{aligned}
\end{align}
Here $\beta$ is a parameter that comes from the hadronization of the dimension-six effective operators, $f_\pi$ is the pion decay constant and $m_p$ and $m_\pi$ are the proton and pion masses, respectively. Finally, $D$ and $F$ are parameters of the baryon number conserving Lagrangian. In Fig.~\ref{fig:protondec} we plot this result as a function of the gauge coupling of $SU(2)_h$ with the band parameterizing our uncertainty on the flavor factor in Eq.~\eqref{eq:vpqrse}. Some phenomenological analyses have been performed in order to constrain this flavor factor~\cite{Lee:2004nz,Lee:2010zzq}. These analyses show that a value of $\left|V_{1111}^e\right|\simeq1$ is compatible with current data. However, in the plot of Fig.~\ref{fig:protondec} we take a conservative lower bound for this factor, $|V_{1111}^e|_{\rm min} = 10^{-5}$. As a result, for a value of $\mu=u=3$ TeV we obtained a bound on the gauge coupling, $g_h< \left[1.1,\,1.3\right]$ with the flavor factor varying from 1 to $10^{-5}$.

\section{Summary and conclusions}
In this talk we have presented an approach to include the flavor structure in the computation of the instanton-generated Green functions. For concreteness, we have done so in the context of the non-universal $SU(2)_l \otimes SU(2)_h \otimes U(1)_Y$ extension of the Standard Model that singularizes the third family. We have have worked in a framework where the instanton Green functions are dominated by the $SU(2)_h$-instanton background. In this framework we have derived the complete set of dimension-six effective operators that reproduce the instanton effects. We have used these operators to derive a bound on the gauge coupling of $SU(2)_h$ for a fixed value of the extended group spontaneous symmetry breaking scale and for different values of the unknown flavor structure of the gauge interactions. This analysis resulted in the bound $g_h< \left[1.1,1.3\right]$. The inclusion of inter-family mixing in the computation was crucial as it allowed us to connect the third family with the other two. This translated in tree-level insertions of the effective operators for the proton decay observables, though suppressed by the non-perturbative exponential factor. Previous attempts to perform this calculation can be found in the literature~\cite{Morrissey:2005uza}. However, flavor interactions were neglected which resulted in higher-loop contributions for the observables considered here. Motivated by the recent experimental bound on $\tau \rightarrow p \mu^+ \mu^-$ by LHCb~\cite{Aaij:2013fia}, the possibility to use $B+L$ violating tau decays in order to constrain the lepton and baryon number violating effective operators was also analyzed in Ref.~\cite{Fuentes-Martin:2014fxa}. However, the authors concluded that an indirect bound on these observables from proton decay set the prospects of observing these decays far beyond the reach of future experiments, something that was already noticed some time ago~\cite{Marciano:1994bg}.

Finally, it should be stressed that even though the results presented in this talk were computed in the framework of the $SU(2)_h$ instantons in the non-universal $SU(2)_l \otimes SU(2)_h \otimes U(1)_Y$ model, they can be easily generalized to other frameworks. For instance, practically the same derivation could be applied to constrain the $SU(2)_l$ gauge coupling from $B+L$ violating proton-proton collisions.

\begin{theacknowledgments}
This research has been supported in part by the Spanish Government, Generalitat Valenciana and ERDF funds from the EU Commission [grants FPA2011-23778, PROMETEOII/2013/007, CSD2007-00042 (Consolider Project CPAN)]. J. F. also acknowledges VLC-CAMPUS for an ``Atracci\'{o} del Talent"  scholarship.   
\end{theacknowledgments}

\bibliographystyle{aipproc}   
\bibliography{references}{}

\end{document}